\documentclass[a4paper,11pt,headings=big,DIV=12]{scrartcl}
\usepackage{amsmath,amssymb}
\usepackage{color,xcolor}
\definecolor{blue}{rgb}{0,0,0.5} 
\usepackage[colorlinks,linkcolor=blue,urlcolor=blue,citecolor=blue]{hyperref}
\usepackage{graphicx}
\addtokomafont{disposition}{\rmfamily\boldmath}
\usepackage[utf8]{inputenc}
\usepackage{slashed}
\usepackage{multirow}
 \usepackage{bbold}
 \usepackage{mathbbol}
 
\usepackage[utf8]{inputenc}
\allowdisplaybreaks
\newcommand{\ONE}{\mathbb{1}}

\newcommand{\vev}[1]{\langle #1 \rangle}

\newcommand{\eps}{\epsilon}

\newcommand{\Tr}{\text{Tr}}

\newcommand{\Rtr}[2]{\Theta}

\newcommand*{\mathcolor}{}
\def\mathcolor#1#{\mathcoloraux{#1}}
\newcommand*{\mathcoloraux}[3]{%
  \protect\leavevmode
  \begingroup
    \color#1{#2}#3%
  \endgroup
}

\begin{document}

\begin{flushright}
\begin{tabular}{l}
UUITP-11/18\\
\end{tabular}
\end{flushright}
\vskip1.5cm

\begin{center}
{\Large\bfseries \boldmath The Conformal Anomaly in bCFT from Momentum Space Perspective}\\[0.8 cm]
{\Large%
Vladimir Prochazka
\\[0.5 cm]
\small
 Department of Physics and Astronomy, Uppsala University,\\
Box 516, SE-75120, Uppsala, Sweden 
} \\[0.5 cm]
\small
E-Mail:
\texttt{\href{mailto:vladimir.prochazka@physics.uu.se}{vladimir.prochazka@physics.uu.se}}
\end{center}

\bigskip

\begin{abstract}
We study the momentum space representation of energy-momentum tensor two-point functions on a space with a planar boundary in $d=3$. We show that non-conservation of momentum in the direction perpendicular to the boundary allows for new phenomena compared to the boundary-less case. Namely we demonstrate how local contact terms arise when the correlators are expanded in the regime where parallel momentum is small compared to the perpendicular one, which corresponds to the near-boundary limit. By exploring two-derivative counterterms involving components of Riemann tensor we identify a finite, scheme-independent part of the two-point function. We then relate this component to the conformal anomaly $c_\partial$ proportional to the boundary curvature $\hat{R}$. In the formalism of this paper $c_\partial$ arises due to integrating out bulk modes coupled to the curved space, which generate local contributions the effective action at the boundary. To calculate the anomaly in specific (free-field) examples, we combine the method of images with Feynman diagrammatic techniques and propose a general methodology for perturbative computations of this type. The framework is tested by computing $c_\partial$ on the explicit example of free scalar with mixed boundary conditions where we find agreement with the literature. 

 \end{abstract}

\newpage 

\tableofcontents

\newpage

\section{Introduction}

The exploration of conformal/trace/Weyl anomaly in even dimensions has a long history \cite{Duff:1993wm}. In the simplest case such anomaly appears when a $2d$ conformal field theory (CFT) without boundary is put on a curved manifold so that the trace of energy-momentum tensor (EMT) receives an anomalous contribution
\begin{eqnarray} \label{eq:2dTraceAnomaly}
\vev{T_\mu^\mu} = c R \; ,
\end{eqnarray}
where $c$ is the central charge of the theory. For a given conformally invariant theory it is possible to compute $c$ unambiguously from the operator product expansion of the EMT. The central charge decreases under RG flow \cite{Zamolodchikov:1986gt} and thus serves as an effective measure of degrees of freedom. Some information about the flow of $c$ can be gained by Fourier transforming the  EMT two-point function and looking at its low momentum behaviour 
\begin{eqnarray} \label{eq:2dMassTT}
\vev{T T} \sim (c_{\text{\tiny{IR}}}- c_{\text{\tiny{UV}}}) p^2 + \mathcal{O}(p^4) \quad. 
\end{eqnarray}
Roughly speaking this means that the IR limit of EMT correlators `knows' about the degrees of freedom integrated out throughout the flow. We will see later on that this concept proves to be relevant in studying boundary anomalies too. \\
The situation becomes drastically different in $3d$, where no conformal anomaly appears.\footnote{The simplest way to see it is that there is no diffeomorphism-invariant, dimension $3$ operators that would contribute to $\vev{T_\mu^\mu}$.} Nevertheless, one might ask whether an anomaly similar to \eqref{eq:2dTraceAnomaly} arises when we insert $2d$ boundary (or defect) in a $3d$ CFT. In this case the symmetries of the theory allow \cite{Schwimmer:2008yh} for a conformal anomaly localised at the boundary
\begin{equation} \label{eq:BoundaryTrace}
\vev{T_\mu^\mu}=  \delta (x_n) \left(c_{\partial} \hat{R} + b \hat{K}_{ij}^2 \right) \; ,
\end{equation}
where $\hat{R}$ is the boundary Ricci scalar and $\hat{K}^2= K_{ij}^2 - \frac{1}{2} K^2 $ is conformally invariant quantity depending on the extrinsic curvature $K_{ij}$. 
The quantity $b$ is a $B-$ type conformal anomaly in the classification of \cite{Deser:1993yx}, whereas  $c_{\partial}$ just like $c$ in \eqref{eq:2dTraceAnomaly} is an $A-$type anomaly corresponding to a topological term.\\
By computing heat kernel coefficients of the Laplacian in the presence of boundaries (cf. \cite{Vassilevich:2003xt} for review of heat kernel computations and their relation to the conformal anomaly) it was shown (see for example \cite{Branson:1990xp} and \cite{Nozaki:2012qd,Fursaev:2015wpa} for more recent discussion) that terms of the type \eqref{eq:BoundaryTrace} appear already in free-field theories. From the geometrical perspective \eqref{eq:BoundaryTrace} can be related to logarithmic corrections to the entanglement entropy \cite{Fursaev:2016inw}, the shape dependence of Casimir effect  \cite{Miao:2017aba} or the anomalous transport in the presence of external magnetic field \cite{Chu:2018ntx}.  More generally, the contribution \eqref{eq:BoundaryTrace} appears whenever a two-dimensional boundary or defect is placed in higher dimensional space.\footnote{The difference is that for $d>3$ other terms involving Weyl tensor appear.} For example in $4d$, the coefficients of \eqref{eq:BoundaryTrace} appear in connection with the Rényi entropy across two-dimensional entangling surface \cite{Bianchi:2015liz}.\\
The anomaly $c_\partial$ was found to depend on boundary conditions and therefore to change under boundary RG flow generated by deforming bCFT through addition of some relevant operators at the boundary (which can be understood as a flow of boundary conditions as we will see in the next section). This together with the analogy to $2d$ central charge \eqref{eq:2dTraceAnomaly} has lead to a conjecture of monotonicity of $c_\partial$ \cite{Nozaki:2012qd} under boundary RG flow and its proof \cite{Jensen:2015swa}. The proof uses the anomaly matching techniques of \cite{Komargodski:2011vj} to find a relation between the change of $c_\partial$ along the flow and the two-point function of the boundary EMT. \\
Recently the relationship between correlators on flat spaces with planar boundaries and boundary central charges has been explored in $4d$ \cite{Herzog:2017xha} and $3d$ \cite{Herzog:2017kkj}. In particular in \cite{Bianchi:2015liz, Herzog:2017kkj} an explicit formula for $b$ anomaly was given by relating the UV divergence in the two-point function of the displacement operator at a planar boundary to a local term $\int_{\partial M} \hat{K}_{ij}^2$ in the effective action. This implies that the $b-$anomaly is related to the \textit{short-distance} behaviour of \textit{boundary} correlators.  In this work we will argue that the $c_\partial$ anomaly appears through the the \textit{near-boundary} behaviour of \textit{bulk} EMT correlators. In this sense the approach of the present work is closer in spirit to the papers \cite{Symanzik:1981wd}, \cite{Diehl1981} and also \cite{McAvity:1992fq}, where an `extra' renormalization appears due to divergences from the near-boundary limit.
We will develop an intuitive understanding in momentum space, where the near-boundary limit corresponds to large perpendicular momentum. The presence of (divergent) gravitational terms \eqref{eq:bulkCterms} in the effective action leads to a \textit{local} contribution to EMT correlators in this limit.\\
The main result of the paper is the formula \eqref{eq:TheRelation} (and more generally \eqref{eq:Main1result}) relating the momentum space two-point function of energy-momentum tensor to the anomaly $c_\partial$. This formula is seen to be consistent with the position space expression recently proposed in \cite{Herzog:2017kkj}. We also show how to use these formulas in explicit examples. Rather than Fourier transforming position space expressions we propose a novel technique introduced in Section \ref{sec:Feynman}, that can be used for direct perturbative computation of the anomaly in momentum space beyond non-interacting theory.
As a side result we derive a consistency condition \eqref{eq:ConsistencyCondCC} and make an observation about the dependence of the anomaly upon the improvement in \eqref{eq:ImprovContr}. 
\\
The paper is structured as follows. In the reminder of the Introduction we will review some essential properties of boundary CFTs. In Section \ref{sec:Momentum} we will discuss the behaviour of EMT correlators in momentum space and the related local counterterms. In Section \ref{sec:TraceBound} we will derive the contribution of counterterms \eqref{eq:bulkCterms} to the trace anomaly. In Section \ref{sec:SchemInv} we will derive a formulas  \eqref{eq:TheRelation}, \eqref{eq:SumRuleMain} for $c_\partial$, by projecting on scheme-independent, traceless component of the EMT two-point function. Finally in Section \ref{sec:Comp} we extend and apply Feynman diagram techniques to compute the anomaly for free scalar with mixed boundary conditions. Appendix \ref{app:C} contains a technical proof of the identity \eqref{eq:TheRelation} and the Appendices \ref{app:A}, \ref{app:B} contain details of the diagrammatic computations for a free scalar field theory with mixed boundary conditions.

\subsection{A brief review of b(oundary)CFT}

In this paper we will consider $3d$ Euclidean CFT on a flat (or slightly curved) space. This theory will be defined by a traceless, conserved energy-momentum tensor (EMT) $T_{\mu \nu}$. We will put this theory on a semi-infinite space $x_3<0$. We will use Cartesian coordinates $x_{\mu}= (x_i,x_3)$, where the Roman indices $i=1,2$ run over the coordinates parallel to the planar boundary placed at $x_3=0$. We will also use the vector notation for vectors parallel to the boundary. \\
We can define boundary-preserving diffeomorphisms $\xi_{\mu}$ as a subset of standard diffeomorphisms
\begin{equation} \label{eq:Bdiff}
\delta g_{\mu \nu} = \partial_{(\mu} \xi_{\nu)}
\end{equation}
satisfying boundary conditions $\xi_3 |_{x_3=0}=0$ and $\partial_{3} \xi_i |_{x_3=0}=0$. Using this one 
can see that the full $3d$ conformal group $SO(4,1)$ is broken down to a subgroup $SO(3,1)$ of boundary-preserving conformal transformations. 
This group will consist of $SO(2)$ rotations, parallel translations, dilations and the two parallel special conformal transformations. Together we have therefore $6$ generators as expected for $SO(3,1)$. For theory to posses this symmetry the EMT needs to satisfy conformal boundary conditions \cite{Cardy:1984bb}
\begin{equation} \label{eq:CFTbc}
T^{i3}|_{x_3=0}=0 \;,
\end{equation}
which means that there is no transport of energy/momentum along the boundary.
Nevertheless, there can still be flow of energy from the boundary corresponding to the perpendicular displacement operator $D$ defined through the boundary condition
\begin{equation}\label{eq:T33BC}
 T_{33}|_{x_3=0}=D(\underline{x}) \quad. 
\end{equation}
In case of non-conformal boundary conditions, the symmetry group is broken down further. In this case the bulk EMT still satisfies the conservation equation $\partial_{\mu} T^{\mu \nu}=0$, but this tensor \textit{doesn't} lead to conserved charges for parallel rotations and translations, since for generic boundary conditions there will be transport of energy/momentum along the boundary. It can be shown that this problem is fixed by defining 
\begin{equation} \label{eq:FullEMT}
T_{\mu \nu} \to T_{\mu \nu} + \delta(x_3) \delta_\mu^i \delta_\nu^j O_{ij}(\underline{x}) \;.
\end{equation}
The new EMT will now define conserved charges for parallel rotations and translations provided 
\begin{equation} \label{eq:TObc}
T^{i3}|_{x_3=0}=\partial_j O^{ij}(\underline{x}) \;.
\end{equation}
By comparison with \eqref{eq:FullEMT} we see that the full conformal group $SO(3,1)$ is broken unless the l.h.s. of \eqref{eq:TObc} vanishes. \\ 
The operator $O_{ij}$ therefore describes a flow of energy at the boundary and it is not conserved (i.e. $\partial_i O_{ij} \neq 0$) in general. 
The equation \eqref{eq:TObc} also reflects the intuitive observation that RG flow on the boundary (r.h.s. of \eqref{eq:TObc}) implies flow of the boundary conditions for bulk operators (l.h.s. of \eqref{eq:TObc}). 
Note that we can always add a conserved $2d$ EMT $t_{ij}$ to $O_{ij}$ without violating the condition \eqref{eq:TObc}. This is equivalent to adding decoupled degrees of freedom at the boundary. The conformal invariance is restored if $O_{ij}$ is conserved and it can be shown that this happens if $O_{ii}=0$ (see \cite{Nakayama:2012ed}). Therefore intuitively we expect to have conformal symmetry whenever boundary degrees of freedom decouple into a $2d$ CFT living at the boundary. In this paper we will consider the extreme case of empty boundary with 
\begin{equation} \label{eq:Assumption}
O_{ij}=0 \; ,
\end{equation}
where only bulk degrees of freedom propagate. \\

\section{A momentum space perspective} \label{sec:Momentum}

This theory will have a well defined correlators for $x_3<0$ so we will analytically extend these functions to whole $3d$ space and define the corresponding Fourier transforms. The following Fourier transform will be of interest to this paper
\begin{equation} \label{eq:TTcorrMom}
\vev{T_{\mu \nu}(\underline{p},p_3)T_{\rho \sigma}(-\underline{p},p_3')}= \int d^dx e^{i px}\int d^d x' e^{i p'x'} \vev{T_{\mu \nu}(\underline{x},x_3)T_{\rho \sigma}(\underline{x}',x_3')} \;.
\end{equation}
Note that because of the presence of boundary $p_3$ is not necessarily equal to $-p_3'$. To define the Fourier transforms we have used dimensional regularization so that the dimension of the parallel subspace is $d-1 = 2 - 2 \epsilon$. The correlators will be defined by coupling the theory to a (bulk) background metric $g_{\mu \nu}$. Had we worked in $d=3$ without a boundary, there would be no logarithmic divergences (or $\epsilon$ poles in dimreg) in the above correlator. One way to understand this is by observing that such divergences would have to correspond to diffeomorphism, scale-invariant local counterterms and there are no such terms in $d=3$. We can however include  diffeomorphism-invariant, scale-dependent ones
\begin{eqnarray} \label{eq:scaleDepCterms}
\int d^3 x \sqrt{g} \Lambda R , \quad  \int d^3 x \sqrt{g} \Lambda^3.
\end{eqnarray}
These would correspond to power divergences and are present in the most generic case.\footnote{In some schemes that preserve scale invariance (for example dimensional regularization) such terms can be set to 0.}

\subsection{Small momentum behaviour of correlators}

Let us consider the Fourier transform of the two point function of exactly marginal (scalar) operator $O$ in $d=3$. In a theory \textit{without} a boundary, the behaviour of the correlator is completely fixed by conformal symmetry \cite{Bzowski:2013sza}. I.e. 
\begin{equation} \label{eq:bulkMomExp}
\vev{O(\underline{p},p_3)O(\underline{p}',p_3')} = C(\underline{p}^2+p_3^2)^{\frac{3}{2}}\delta^{(2)}(\underline{p}+\underline{p}') \delta(p_3+p_3') \;.
\end{equation}
The above behaviour cannot correspond to any local contact term.\\
Adding a planar boundary at $x_3=0$ breaks the translational symmetry in the perpendicular direction, which makes the momentum space structure much richer. We expect two-point functions to be distributionally supported at $p_3 \neq - p_3'$ in this case. We will see later in explicit examples how this new dependence arises, but for now we would just like to restrict to some general arguments using dimensional analysis. Throughout this paper, we would like to focus on special kinematics with $p_3>0, p_3'=0$. Such specific kinematics should be sensitive to boundary effects since it can be thought about as an offset to the $p_3=-p_3'$ conserved limit. Physically the relevant amplitude should correspond to a process where the incoming state gets absorbed by the boundary. We will further consider an 'effective IR' limit $p_3 \gg |\underline{p}|$. This limit is special, since it should probe the physics of the near boundary region $|\underline{x}| \gg |x_3|, |x_3'|$. Using dimensional analysis (or scale invariance) one can deduce the small $\underline{p}^2$ expansion of the correlator \footnote{More generally , the dimensional analysis restricts the low $\underline{p}^2$ behaviour to be of the form $\left(f_1(\frac{\underline{p}^2}{p_3^2})\underline{p}^2 + f_2(\frac{\underline{p}^2}{p_3^2})p_3^2 \right)\delta^{(2)} (\underline{p}+\underline{p}')$. In the absence of divergences associated with the boundary limit, the small $\underline{p}^2$, large $p_3$ behaviour should be smooth. From this we expect the functions $f_{1,2}$ to asymptote to  finite constants $C_{1,2}$ for small values of the argument.} 
\begin{equation} \label{eq:LowMomOO}
\vev{O(\underline{p},p_3)O(\underline{p}',0)} = \left( C_1\underline{p}^2+C_2 p_3^2 + \mathcal{O} \left(\frac{\underline{p}^4}{p_3^2}\right) \right) \delta^{(2)} (\underline{p}+\underline{p}')  \; ,
\end{equation}
where the constants $C_{1,2}$ depend on the boundary conditions and aren't necessarily equal to each other since the global rotational symmetry $SO(3)$ is now broken down to $SO(2)$.  Above we have assumed that the correlator is well behaved for small $|\underline{p}|$ due to $p_3>0$ acting as an effective IR cutoff. Furthermore we have assumed that no divergences arise from performing the Fourier integral over boundary region $x_3=x_3'=0$ (we will consider the divergent case below).  The absence of extra delta function yields new \textit{local} terms $\underline{p}^2,  p_3^2$ in the low $\underline{p}^2$ expansion. Indeed, one readily verifies that for example 
\begin{equation} \label{eq:LocalCterms}
\delta(x_3)\partial_{i}^2 \delta(x-x')  \stackrel{F.T}{\to} - \underline{p}^2 \delta^{(2)} (\underline{p}+\underline{p}') \quad ,
\end{equation}
which comes from a local two-derivative term at the boundary. Higher order contributions to \eqref{eq:LowMomOO} will be non-local terms like $\frac{\underline{p}^4}{p_3^2}$ etc. \\
Additionally, if the correlator \eqref{eq:LowMomOO} contains divergences from integrating over the boundary we can have an extra non-local
contribution
\begin{equation}
\left[ b_1 \underline{p}^2 \ln(\frac{p_3}{\mu})+ b_2 p_3^2 \ln(\frac{p_3}{\mu})  \right] \delta^{(2)}(\underline{p}+\underline{p}')  \quad ,
\end{equation}
which is related to the appearance of $B-$type conformal anomaly. Let us comment that above we didn't describe the \textit{full} correlator in the momentum space. In general there will also be contributions proportional to $\delta(p_3)$ supported at $p_3=p_3'=0$. For example the scale invariance allows the following term
\begin{equation} \label{eq:ConsMomContribution}
C|\underline{p}|^3\delta^{(2)}(\underline{p}+\underline{p}') \delta(p_3) \in \vev{O(\underline{p},p_3)O(\underline{p}',0)} \;,
\end{equation}
which includes the bulk contribution \eqref{eq:bulkMomExp}.\\
Before we proceed, let us offer some intuitive understanding of the small momentum expansion \eqref{eq:LowMomOO}. The component $p_3$ plays a role of the mass parameter in the two-point functions. The intuitive picture is that since there are no propagating boundary degrees of freedom\footnote{More precisely, there are no modes propagating \textit{along} the boundary owing to \eqref{eq:Assumption}. Since $T_{33} \neq  0$, there will still be energy/momentum transfer \textit{perpendicular} to the boundary. However due to \eqref{eq:CFTbc}, this perpendicular momentum flow doesn't produce any modes along the boundary.}   the bulk fields should get 'integrated out' for large $p_3$ leaving the boundary theory empty up to some local terms required by anomaly matching. It is these local terms, which should give us the boundary anomalies. This is analogical to a $2d$ QFT theory where all degrees of freedom are gapped and IR theory consists of pure contact term correlators \cite{Bonora:2015odi}. \\
In the next subsection we will describe how such local terms of the form \eqref{eq:LocalCterms} arise in the two- point function \eqref{eq:TTcorrMom}.

\subsection{Boundary counterterms and new divergences} \label{sec:BoundaryCT}
In this subsection we proceed to discuss the contribution of local boundary diffeomorphism invariant counterterms to the correlator \eqref{eq:TTcorrMom}. Before we start we need to define a consistent variation principle for functional derivatives w.r.t. $g_{\mu \nu}$, which will follow from setting the boundary conditions for the metric. To be consistent with boundary diffeomorphisms \eqref{eq:Bdiff} we take
\begin{eqnarray} \label{eq:MetricBC} \notag
\partial_3 g_{ij}|_{x_3=0}&=& 0 \\
 g_{i3}|_{x_3=0} &=& 0 \quad.  
\end{eqnarray}
Another way to understand the above conditions is that we want to extend the metric to the unphysical region $x_3>0$ to apply the method of images. In particular we will want the $g_{ij}$ to be symmetric (parity-even) w.r.t. reflections $x_3 \to -x_3$. All the parity-odd components have to vanish at the boundary, which is equivalent to \eqref{eq:MetricBC}. \\
The variational principle should reflect these conditions, which can be achieved by method of images. In particular the first condition in \eqref{eq:MetricBC} corresponds to a Neumann boundary condition on the allowed contact terms of $T_{ij}$ correlators. This means for example that all the contact terms in the point function of $T_{ij}$ should be replaced
\begin{equation} \label{eq:BoundVariation}
\delta^{(3)}(x-y) \to \delta^{(3)}(x-x') + \delta^{(3)}(\tilde{x}-x') \quad,
\end{equation}
where  $\tilde{x}= (\underline{x},-x_3)$ is the position of image in the unphysical region. Such delta function has the desired property that it is equal to $\delta^{(3)}(x-y)$ in the bulk and satisfies the Neumann boundary condition. With this in mind we can proceed with the discussion. \\
We start by writing down the two admissible local counterterms formed out of Riemann tensor components
\begin{equation} \label{eq:bulkCterms}
\int d^3 x \delta(x_3) \sqrt{g} R ,\quad  \int d^3 x \delta(x_3) \sqrt{g} g^{ij}R_{i3j}^3 ,
\end{equation}
where $\delta (x_3)$ is the normal delta function localizing to the boundary surface. Both of these terms have two derivatives and are invariant with respect to the boundary-preserving (parallel) diffeomorphisms. Furthermore, these counterterms are scale invariant.\\
In addition, we can write further boundary terms depending on the extrinsic curvature
\begin{equation} \label{eq:boundCterms}
\int d^2 \underline{x}\sqrt{\hat{g}}{\hat{K}_{ij}}^2, \quad  \int d^2 \underline{x} \sqrt{\hat{g}}K^2,  \quad \int d^2 \underline{x}\sqrt{\hat{g}} \partial_3 K ,
\end{equation}
where $\hat{K}_{ij}=K_{ij}- \frac{1}{(d-1)} \hat{g}_{ij} K^2$ and $\hat{R}$ is the Ricci scalar of the boundary metric $\hat{g}_{ij}$.\footnote{In a suitable coordinate system (Gaussian coordinates) we can take $\hat{g}_{ij}= \lim_{x_3 \to 0} g_{ij}$.} The extrinsic curvature $K_{ij}$ depends only on normal derivatives of $g_{ij}$ and $g_{i 3}$ at the boundary so the first two terms above are not to be included when imposing the boundary conditions \eqref{eq:MetricBC}. The  coefficients of these terms are fixed by the short-distance behaviour of correlators of pure boundary operators.
For example the first term was shown \cite{Herzog:2017kkj} to be related to the short-distance divergences of the two-point function of the displacement operator \eqref{eq:T33BC}. More specifically it was shown that $\vev{D(\underline{x})D(0)}$ contains divergences proportional to $\underline{\partial}^4 \delta^{(d-1)}(\underline{x})$. This contact term has vanishing Fourier transform if we take $p_3>0$ and hence doesn't appear in correlators of the type \eqref{eq:LowMomOO} studied in this paper.\footnote{The fact that contact terms of $T_{33}$ in the bulk can be studied separately from those of $D$ follows from the observation that the respective sources should be as independent quantities with the condition \eqref{eq:T33BC} valid only at non-coincident points.}  \\
In what follows we will focus solely on \eqref{eq:bulkCterms}.
Apart from \eqref{eq:bulkCterms}, \eqref{eq:boundCterms} there are also other admissible boundary counterterms, but these are either total derivatives or not scale-invariant so we will not discuss them here. \\
The counterterms in \eqref{eq:bulkCterms} might also have divergent coefficients in general. These divergences don't come from the short-distance limit as usual but rather from the near-boundary behaviour.  For example if the theory has an operator $O$ of dimension $d-2$, the correlator of the trace of EMT (which is pure contact term in CFT) includes the following contribution
\begin{equation} \label{eq:SemiLocalContact}
 \vev{O}\partial^2 \delta^{(3)}(x-x') \in \langle T_\mu^\mu(x)T_\rho^\rho(x') \rangle \;.
\end{equation}
In a CFT without a boundary such terms vanish since $\vev{O}=0$, but in the presence of boundary we have $\vev{O} \propto \frac{1}{|x_3|^{d-2}}$. Hence for bCFT such contact term becomes semi-local and acquires non-trivial Fourier transform. For example a following contribution could arise
\begin{equation} \label{eq:SemiLocalContactMom}
 \vev{O}\delta(x_3-x_3')\partial_i^2 \delta^{(d-1)}(\underline{x}-\underline{x'})
 \stackrel{F.T.}{\to}\frac{1}{\eps}\underline {p}^2  p_3^{-2 \eps} \delta^{(d-1)}(\underline{p}+\underline{p}')\propto \underline {p}^2 (\frac{1}{\epsilon} + \mathcal{O}(\epsilon^0)) \;.
\end{equation}
Similarly the two-point function of EMT usually contains a term proportional to
\begin{eqnarray}
\delta^{(d)}(x-x')\vev{T_{\mu \nu}} \sim \delta(x_3-x_3')\delta^{(2-2 \eps)}(\underline{x}-\underline{x'}) \frac{1}{|x_3|^{3 - 2 \eps}}  \stackrel{F.T.}{\to}p_3^2\left(\frac{1}{\epsilon} + \mathcal{O}(\epsilon^0)\right)\delta^{(d-1)}(\underline{p}+\underline{p}') \;.
\end{eqnarray}
One possible source of contact terms of the type \eqref{eq:SemiLocalContact} are for example equations of motion. As we will see in the explicit examples (e.g. in Appendix \ref{app:B}) new divergences will appear when evaluating loops for Feynman diagrams with non-conserved $p_3$. Such divergences should be removed by adding covariant counterterms localised on the boundary \cite{Symanzik:1981wd}. Even if the corresponding correlator doesn't have divergences, the choice of counterterm coefficient represents a contact term ambiguity. Physical quantities shouldn't involve such ambiguity so it important to understand, which correlators are invariant.  \\
To define renormalized correlators in dimensional regularization we will apply the quantum action principle \cite{Brown:1980qq}. The two-point functions can be defined through second derivative of the partition function
\begin{equation} \label{eq:QAPTT}
\vev{ T_{\mu \nu}(x) T_{\rho \sigma}(x')} = -4 \frac{\delta^2 W}{\delta g_{\mu \nu}(x)\delta g_{\rho \sigma}(x')} \big |_{g_{\mu \nu} \to \delta_{\mu \nu}}  \quad.
\end{equation}
where the effective action (generating functional of connected Green's functions) was defined $W= -\ln \int \mathcal{D}(\phi) e^{-S}$ and the variational rule \eqref{eq:BoundVariation} is assumed. By the quantum action principle this correlator should be finite. More precisely we can rewrite \eqref{eq:QAPTT} as 
\begin{equation} \label{eq:TTCorrDef}
\vev{ T_{\mu \nu}(x) T_{\rho \sigma}(x')}= -4\vev{ \frac{\delta S}{\delta g_{\mu}(x)}  \frac{\delta S}{\delta g_{\rho \sigma}(x')} } \big |_{g_{\mu \nu} \to \delta_{\mu \nu}}+ 4\vev{ \frac{\delta^2 S}{\delta g_{\mu \nu}(x)\delta g_{\rho \sigma}(x')}}\big |_{g_{\mu \nu}\to \delta_{\mu \nu}} \quad ,
\end{equation}
where the fist term on r.h.s. of \eqref{eq:TTCorrDef} can be thought of as the bare part, whereas the second term includes contact terms necessary to renormalize the correlator.\footnote{Note that this definition only includes \textit{connected} Green's functions. }  \\
Let us now check how \eqref{eq:bulkCterms} contribute to some particular two-point functions of $T_{\mu \nu}$ by writing a counterterm action
\begin{eqnarray} \label{eq:Scounter}
\delta S= \mu^{-2 \eps}\int d^d x \delta(x_3)\sqrt{g}\left(c_1 R + c_2 g^{ij}R_{i3j}^3\right) \; ,
\end{eqnarray}
where an arbitrary mass scale $\mu$ was included to preserve the scale-invariance. \\
Expanding these terms to quadratic order in the metric and taking the variations we can find their contribution to various correlators of energy-momentum tensor. In particular we have
\begin{eqnarray} \label{eq:RdoubleVariation}
\delta \vev{T_{ij} T_{kl}} = 4 \frac{\delta^2}{\delta g_{ij}(x)\delta g_{kl}(x')} \delta S \; |_{g_{\mu \nu}= \delta_{\mu \nu}}\stackrel{F.T.}{\to} c_1 \underline{p}^2 \delta^{(d-1)}(\underline{p}+ \underline{p}') \times \mathcal{O}(d-3) + c_2 \times 0 \quad ,
\end{eqnarray}
where we omitted the (non-zero) part proportional to $p_3^2, p_3'^2$, since it is not relevant for our purposes.\footnote{Divergences proportional to $p_3^2$ are renormalized through contact terms of the type $\delta(x-x')\vev{T_{\mu \nu}}$ (we include some discussion of this in Appendix \ref{app:B}).   
We can always avoid these terms by projecting on $\underline{p}^2$ via momentum derivatives.} In \eqref{eq:RdoubleVariation} we have emphasized that $c_1$ gives evanescent contribution near $d=3$ and $c_2$ vanishes identically. Since \eqref{eq:bulkCterms} are the only counterterms yielding $\underline{p}^2$ terms, we immediately deduce two things. First, it follows that in $d=3$ the two-point functions of purely parallel components of $T_{\mu \nu}$ should not include any divergences proportional to $\underline{p}^2$. The second implication is that this component is invariant w.r.t. scheme change induced by a finite shift in $c_1, c_2$ in \eqref{eq:Scounter}. This is reminiscent of what happens with $\int \sqrt{g}R$ counterterm in $2d$ and is related to the topological character of $\int_{\partial M} \hat{R}$. \\ 
On the other hand the correlator of perpendicular components receives contribution from \eqref{eq:Scounter}
\begin{eqnarray}\label{eq:RndoubleVariation}
\delta \vev{T_{33} T_{33}} = 4 \frac{\delta^2}{\delta g_{33}(x)\delta g_{33}(x')} \delta S\; |_{g_{\mu \nu}= \delta_{\mu \nu}}\stackrel{F.T.}{\to} -\left(c_1+\frac{1}{2}c_2 \right)\underline{p}^2\delta^{(d-1)}(\underline{p}+ \underline{p}') \;.
\end{eqnarray}
We see that the $\underline{p}^2$ divergent part of the coefficient $c_2$ is fully determined from the $\underline{p}^2$ divergence of $\vev{T_{33}T_{33}}$ and the knowledge of $c_1$. As we will see in the next section conformal invariance implies that the $\underline{p}^2$ divergent part of $\vev{T_{33}T_{33}}$ vanishes.\footnote{This is of course consistent with traceless character of $T_\mu^\mu$, which implies that $T_{33}=-T_{ii}$. However, this identity only holds on shell and we saw above that equations of motion can produce non-trivial divergences.} Nevertheless, the (finite) $\underline{p}^2$ part of $\vev{T_{33}T_{33}}$ is still arbitrary and dependent on the choice of $c_1, c_2$. \\
The correlators involving $T_{3i}$ receive no $\underline{p}^2$ contributions from \eqref{eq:Scounter} or any other counterterm due to boundary conditions \eqref{eq:MetricBC} and the related variation rule \eqref{eq:BoundVariation}. These correlators should not have any divergences from the near-boundary limit due to the boundary condition \eqref{eq:CFTbc}.

\subsection{The trace anomaly} \label{sec:TraceBound}

We would now like to discuss how \eqref{eq:bulkCterms} contribute to the trace anomaly. In order to that we will extend the standard dimensional regularization arguments \cite{Duff:1977ay}  to the case of a theory with boundary. In this approach the (bare) effective action is conformally invariant, but contains divergences  $W= \frac{c}{\epsilon} + \dots$. These divergences need to be subtracted using suitable local counterterms proportional to $\frac{\mu^{-2\eps}}{\eps}$, which break the conformal invariance and the trace anomaly arises. In the case of $d=3-2\eps$, there is no bulk anomaly so the divergent part of $W$ should be localised on the boundary
\begin{equation} \label{eq:Wbound}
W_{an}  \stackrel{\tiny{d \to 3}}{\approx}  \frac{1}{2 \eps} \int d^d x \sqrt{g} \delta(x_3) \left(c_\partial R + b_n g^{ij}R_{i3j}^3\right) + \dots \quad.
\end{equation}
We expect this term to arise from near boundary behaviour of the bulk effective action integral. To guess this behaviour we will isolate the part of the action integral coming from an infinitesimal strip of width $\delta$ near the boundary. Using covariance w.r.t. boundary diffeomorphisms and scale invariance we expect the relevant contribution to action to be of the form
\begin{equation} \label{eq:Wbulk}
 \int_{-\delta}^{0}dx_3 \int d^{d-1} \underline{x} \sqrt{g}  |x_3|^{2 \eps -1}  \left(c_\partial R + b_n g^{ij}R_{i3j}^3\right) \quad. 
\end{equation} 
The bulk action is finite (there are no bulk anomalies in odd dimensions) so both $c_\partial, b_n$ should be finite and well defined quantities.
Using the identity $\frac{1}{|x|^{1-2\eps}} \stackrel{\eps \to 0}{\approx} \frac{1}{2 \eps} \delta(x) + \dots$ we find that indeed \eqref{eq:Wbulk} reduces to \eqref{eq:Wbound} in the $d \to 3$ limit. The divergence in \eqref{eq:Wbulk} comes from the near boundary limit $x_3 \to 0$, which is consistent with the discussion in the Section \ref{sec:BoundaryCT}. Had we used the cutoff regularization $x_3 < \frac{1}{\Lambda}$ instead of dim-reg, the divergence would reappear as $\log \Lambda$. \\
From \eqref{eq:Wbulk} we see that even though the bulk effective action in non-local its contribution to the boundary part \eqref{eq:Wbound} is \textit{local} (polynomial in $\underline{p}^2$). This doesn't mean that the boundary part of the effective action is local in general.  There will be non-local contributions corresponding to the sources of operators living at the boundary. For example if we took $O_{ij} \neq 0$ in \eqref{eq:FullEMT}, the effective action would contain non-local terms depending on $g_{ij}$ at the boundary. \\
By taking two metric variations of \eqref{eq:Wbound} and Fourier transforming the result we can identify the anomaly contribution to the two-point function of $T_{ij}$\footnote{When performing the metric variation and Fourier transform we have to be careful with factors of $2$. First we pick up a factor of $2$ from the `image' delta function (cf. \eqref{eq:BoundVariation}) and we also pick up a factor of $2$ from integrating $\delta(x_3)$ over the unphysical region $x_3>0$.} 
\begin{eqnarray} \label{eq:Tan}
 \left(-4 \frac{\delta^2 W_{an}}{\delta g_{ij}(x)\delta g_{kl}(x')}|_{g_{\mu \nu} \delta_{\mu \nu}}\right) \stackrel{F.T.}{\to}\left(\frac{4}{\eps}\right) c_\partial  T_{ijkl} + \dots \; ,
\end{eqnarray}
where
\begin{equation}\label{eq:Tan1}
T_{ijkl}= \underline{p}^2 \left[P_{ij}P_{kl}-\frac{1}{2}(P_{ik}P_{lj}+P_{il}P_{jk})\right] 
\end{equation}
and $P_{ij}= \frac{1}{\underline{p}^2} (p_i p_j - \delta_{ij} \underline{p}^2)$. The dots stand for the $p_3^2$ term and finite contributions generated from the regular part of $W$.  The operator $T_{ijkl}$ is transverse, local in $p_i$ and evanescent close to $d=3$ as expected already from \eqref{eq:RdoubleVariation}. In fact, tracing \eqref{eq:Tan} we get
\begin{equation}
-8 c_\partial P_{kl} 
\end{equation}
which is analogical to the contact term appearing in the $2d$ trace anomaly.\footnote{In fact by writing the boundary metric in the conformal gauge $\hat{g}_{ij}=e^{-2 \hat{\tau}} \delta_{ij}$ we find that the finite part of $W_{an}$ contains the term $\int d^2 \underline{x} \hat{\tau} \Box \hat{\tau}$, which is exactly the anomaly-induced (dilaton) effective action in a $2d$ CFT.} 
 Even though the correlators of energy-momentum tensor are finite, the divergent effective action is renormalized via defining $W_{fin}= W+ S_{ct}$, where the counterterm action $S_{ct}$ of the form \eqref{eq:Scounter} has
\begin{equation} \label{eq:CdnAnomalyDef}
c_1 = - \frac{1}{2 \eps} c_{\partial} ; \quad c_2 = - \frac{1}{2 \eps} b_n \quad.
\end{equation}
These counterterms break scale invariance through the arbitrary scale $\mu$ and the trace anomaly appears \cite{Birrell:1982ix}
\begin{equation} \label{eq:Tmm1}
   \int d^d x \sqrt{g} \vev{T_\mu^\mu} =  -\mu\frac{\partial W_{fin}}{\partial \mu}  =- \mu  \frac{\partial S_{ct}}{\partial \mu} = - \int d^d x \sqrt{g} \delta(x_3) \left(c_\partial R + b_n g^{ij}R_{i3j}^3\right) \;.
\end{equation}
We should stress that in \eqref{eq:Tmm1} we only included \textit{bulk} contribution to the trace anomaly. In general one should also include the terms \eqref{eq:boundCterms} in $W_{an}$.
 In particular the first term is important for the $b-$anomaly as was discussed in \cite{Herzog:2017kkj}.\footnote{The second term in \eqref{eq:boundCterms} that is forbidden by Wess-Zumino consistency conditions can always be cancelled by a suitable choice of Robin boundary conditions, whereas the last term can be removed by a choice of scheme.}
 \\
To understand the anomalies \eqref{eq:Tmm1} we can use Gauss-Codazzi relations
\begin{eqnarray}\notag
R &=& \hat{R} + 2 a_i^2 - K^2 - K_{ij}^2 + \dots \\ \label{eq:GCrels}
g^{ij}R_{i3j}^3 &=& - K_{ij}^2 +   a_i^2 + \dots \quad ,
\end{eqnarray}
where the dots stand for total derivative terms and $a_i= (\nabla_3 n)_i = \Gamma_{3i}^3 \sim \partial_i g_{33}$ is the acceleration vector. Note that unlike the extrinsic curvature , the  $a_i^2$ terms don't vanish under boundary conditions \eqref{eq:MetricBC}. However, such term is ruled out in a confromal theory by Wess-Zumino consistency conditions\footnote{It should be noted that Wess-Zumino constraints are only valid for strictly conformal theories with vanishing $T_\mu^\mu$. In some cases (like free scalar theory for example) conformal boundary conditions must be imposed by adding suitable improvement terms on the boundary.} \cite{Schwimmer:2008yh} and hence we obtain a constraint
\begin{eqnarray}\label{eq:ConsistencyCondCC}
2 c_\partial + b_n = 0 \quad.
\end{eqnarray}
The consistency condition \eqref{eq:ConsistencyCondCC} implies that the contribution of the intrinsic curvature to the anomaly always appears in the combination
\begin{equation}
 R- 2 R_{i3j}^3 \; 
\end{equation}
which is also confirmed by the free field computations of \cite{Vassilevich:2003xt}. 

By combining this condition with \eqref{eq:CdnAnomalyDef} and \eqref{eq:RndoubleVariation} we conclude that $\vev{T_{33}T_{33}}$ has no $\underline{p}^2$ divergences in a conformally invariant theory. However the finite $\underline{p}^2$ part is not fixed by any symmetry principle (see the discussion under \eqref{eq:RndoubleVariation}) so this component cannot be proportional to a physical quantity. 
Since there is no invariant counterterm proportional to normal derivatives of $g_{33}$ we conclude that also no $p_3^2$ divergences can appear. Altogether we arrive at the result that the bulk correlator $\vev{T_{33}T_{33}}$ is finite and scheme-dependent.
In Appendix \ref{app:B}, we verified that the relation \eqref{eq:ConsistencyCondCC} is satisfied for conformally coupled free scalar with mixed boundary conditions.\\
Finally, neglecting the other terms involving extrinsic curvature and total derivatives we get 
\begin{eqnarray} \label{eq:TraceAnBoundary}
 \vev{T_\mu^\mu}= - \delta(x_3) c_\partial \hat{R} \quad,
\end{eqnarray}
which shows that the Ricci scalar anomaly is determined directly from the $\frac{1}{\eps}$ pole of $R$ in the effective action. In the next section we will show how to relate this coefficient to the correlators of bulk EMT. \\

\subsection{Searching for a physical quantity} \label{sec:SchemInv}
We would now like to look for a finite, scheme-independent quantities to identify a suitable candidate for $c_\partial$. By scheme we mean the choice of finite coefficients of counterterms like \eqref{eq:bulkCterms}. \\
From discussion of the previous section (cf. \eqref{eq:RdoubleVariation}) we see that the purely parallel component $\vev{T_{ij}(\underline{p},p_3)T_{kl}(-\underline{p},p_3')}$ receives no $\underline{p}^2$ corrections from \eqref{eq:bulkCterms} so we will project on its $\underline{p}^2$ part. The resulting quantity still receives correction from the first term in \eqref{eq:scaleDepCterms}. To remove this ambiguity we can choose a configuration of external momenta which explicitly violates momentum conservation with $p_3>0$ and $p_3'=0$. Thus we see that a desired scheme-invariant quantity should be some combination of terms of the type
\begin{eqnarray} \label{eq:Combinations}
\frac{\partial}{\partial \underline{p}^2} \vev{T_{ij}(\underline{p},p_3)T_{kl}(-\underline{p},0)} \; ,
\end{eqnarray}
where $\frac{\partial}{\partial \underline{p}^2}$ stands for some second order differential operator in $p_i$ that projects on the right kinematic structure. The remainder of this section is devoted to finding the correct combination proportional to $c_\partial$ and determining the constant of proportionality. \\
 A detailed argument identifying the anomaly is offered in the Appendix \ref{app:C}. Here we would just like to summarize its main points. As seen in Section \ref{sec:TraceBound} the part of the two-point function $\vev{T_{ij}T_{kl}}$ related to the anomaly (cf. \eqref{eq:Tan}, \eqref{eq:Tan1}) is transverse.  It turns out that the non-transverse terms don't contribute to the two-point function of the traceless component  
\begin{eqnarray} \label{eq:TracelessT}
\tilde{T}_{ij}= T_{ij} - \frac{\delta_{ij}}{(d-1)}T_{kk} \; ,
\end{eqnarray}
which is therefore a good candidate to extract the anomaly from.
The expansion \eqref{eq:LowMomOO} applied to the correlator of $\tilde{T}_{ij}$ has the form
\begin{equation} \label{eq:TilTil1}
    \vev{\tilde{T}_{ij}(\underline{p},p_3)\tilde{T}_{kl}(\underline{p'},0)}= (\tilde{A}_{ijkl}+ \tilde{B}_{ijkl}p_3^2)\delta^{(d-1)}(\underline{p}+\underline{p}')+ \mathcal{O}\left(\frac{\underline{p}^4}{p_3^2}\right) \quad , 
\end{equation}
where $\tilde{A}_{ijkl}, \tilde{B}_{ijkl}$ are traceless $SO(2)-$covariant quantities and $\tilde{A}_{ijkl}$ quadratic in $\underline{p}$. By the above reasoning we get
\begin{equation} \label{eq:Main1result}
\tilde{A}_{ijkl} \sim c_\partial \;.
\end{equation}
 To simplify the computation and comparison with \eqref{eq:Tan} it is convenient to contract with $p_i$, which leads to the final result
\begin{equation} \label{eq:TheRelation}
p_i \tilde{A}_{ijkl}= -4 c_\partial pj \left(p_k p_l - \frac{1}{(d-1)} \delta_{kl} \underline{p}^2\right) \;.
\end{equation}
Additionally if there exists a bulk (scalar) operator $O$ of dimension $1$ in the theory, the formula for $c_\partial$ receives a linear shift under the improvement of the form \eqref{eq:ImprovementGeneral}. More concretely by combining \eqref{eq:MFImprov} and \eqref{eq:Wbulk} with $\vev{O}= \frac{C_O}{|x_3|^{(d-2)}}$ this shift can be evaluated explicitly
\begin{equation}\label{eq:ImprovContr}
\delta c_\partial = \frac{1}{2} \xi C_O \;.
\end{equation}
The linear dependence of $c_\partial$ on the value of the improvement is surprising given that the bulk central charges in $4d$ typically don't depend on improvements. The dependence of $c_\partial$ on the improvement was already observed for the free scalar in \cite{Fursaev:2016inw}, here we generalize it to a generic improvement \eqref{eq:ImprovementGeneral}. Practically speaking it is nevertheless necessary to choose a specific $\xi_c$ in the above formula to enforce the conformal symmetry via $T_\mu^\mu=0$, otherwise some other anomalies appear (cf. the discussion below \eqref{eq:GCrels}). \\
Coming back to the formula \eqref{eq:TheRelation} we notice that it allows for direct evaluation of the anomaly from the knowledge of $\vev{ \tilde{T}_{ij} \tilde{T}_{kl}}$ in momentum space. In the next section we will propose a novel perturbative method to compute the correlators directly in momentum space.\\
In general one can get $c_\partial$ directly from $\vev{ \tilde{T}_{ij} \tilde{T}_{kl}}$ for example by taking momentum derivatives of a suitable contraction of \eqref{eq:TilATen}
\begin{eqnarray} \label{eq:kappa1} \notag
-4 c_\partial &=& \frac{1}{8} \frac{\partial}{\partial p_k}\frac{\partial^2}{\partial p_l^2} p_i \vev{\tilde{T}_{ij}(\underline{p},p_3) \tilde{T}_{jk}(-\underline{p},0)} \big |_{\underline{p} \to 0} \\
&=& \frac{1}{8}\left( \frac{\partial^2}{\partial p_l^2}  \vev{\tilde{T}_{ij}(\underline{p},p_3) \tilde{T}_{ij}(\underline{p'},0)}+ 2  \frac{\partial^2}{\partial p_i \partial p_k} \vev{\tilde{T}_{ij}(\underline{p},p_3) \tilde{T}_{jk}(-\underline{p},0)}  \right) \big |_{\underline{p} \to 0} \;.
\end{eqnarray}
 This result can also be cast into a sum-rule relation in the position space
 \begin{equation} \label{eq:SumRuleMain}
c_{\partial}= \frac{1}{32} \int d^2 \underline{x} dx_3 dx_3' e^{ip_3 x_3}\left(\underline{x}^2  \vev{\tilde{T}_{ij}(\underline{x},x_3)\tilde{T}_{ij}(0,x_3')} + 2 x_i x_j \vev{\tilde{T}_{ik}(\underline{x},x_3)\tilde{T}_{jk}(0,x_3')} \right) \; ,
 \end{equation}
 where the operator $\tilde{T}$ was defined in \eqref{eq:TracelessT}. \\
 At last we would like to remark on the consistency between \eqref{eq:Main1result} and the position space formula for $c_\partial$ proposed in \cite{Herzog:2017kkj} . In \cite{Herzog:2017kkj} it was advocated that $c_\partial$ can be written as linear combination of the coefficients $\alpha(1), \epsilon(1)$ which describe the boundary limit of $\vev{T_{33}T_{33}}$ and $\vev{ \tilde{T}_{ij} \tilde{T}_{kl}}$ in position space. Based on  \eqref{eq:IRconserved} and \eqref{eq:Main1result} one could be 
tempted to conclude that $c_\partial$ only depends on $\epsilon(1)$. This is complicated by the fact that
 the  amplitude \eqref{eq:TilTil1} is evaluated at special kinematics $p_3'=0, p_3>0$ and thus excludes some information from the original position space correlator. More concretely it excludes the $p_3=0$ contribution of the type \eqref{eq:ConsMomContribution}, which comprises the conserved $d=3$ bulk part.
It is therefore quite reasonable to expect the position space expression for the anomaly to be some combination of coefficients $\epsilon$ and $\alpha$ independent of the bulk part.
  In the language of \cite{McAvity:1993ue}, the bulk part corresponds to the $v \to 0$ limit of the functions $\alpha(v), \epsilon(v)$. This component is constrained by conformal Ward identities that relate the two coefficients
 \begin{equation*}
 \epsilon(0)= \frac{3}{4} \alpha(0) \;.
 \end{equation*}
Thus we see that the combination
\begin{equation} \label{eq:HHJformula}
\epsilon(1)- \frac{3}{4} \alpha(1)
\end{equation}
is completely independent of the bulk contribution. Indeed, \eqref{eq:HHJformula} is the combination that appears in \cite{Herzog:2017kkj}.\footnote{The third term in \cite{Herzog:2017kkj} comes from a decoupled boundary energy-momentum tensor which we set to zero by the condition \eqref{eq:Assumption}.}

\section{Computing the anomaly} \label{sec:Comp}

\subsection{Free scalar}
 
In this section we will illustrate the main ideas of this paper on the explicit example of free scalar field with mixed boundary conditions. In the bulk, this theory is described by the following action
\begin{equation} \label{eq:S0scalar}
S_0= \frac{1}{2}\int_{x_3<0} d^d x \sqrt{g} \left((\partial \phi^A)^2+ \xi R (\phi^{A})^2 \right) \quad ,
\end{equation}
where we have summed over $N$ non-interacting scalar fields. 
The conserved bulk EMT for \eqref{eq:S0scalar} reads
\begin{equation}\label{eq:scalarEMT}
T_{\mu \nu}=  \frac{-2}{\sqrt{g}}\frac{\delta}{\delta g_{\mu \nu}(x)} S_0 \big |_{g_{\mu \nu}= \delta_{\mu \nu}}= \partial_\mu \phi^A \partial_\nu \phi^A - \frac{1}{2}\delta_{\mu \nu} (\partial \phi^A)^2 + \xi (\delta_{\mu \nu}\Box - \partial_{\mu} \partial_{\nu}) (\phi^A)^2 \quad.
\end{equation}
For $T_{\mu \nu}$ to be traceless we would need to take $\xi_c= \frac{(d-2)}{4(d-1)}$, but we will use generic $\xi$ in the following to achieve more generality. \\
We can express the mixed boundary condition by using projectors $\Pi_-$ and $\Pi_+$ which project onto subspaces satisfying Dirichlet and Neumann conditions respectively. More concretely by expressing $\phi^A$ as a column vector and projectors as matrices the boundary conditions read \cite{Vassilevich:2003xt}
\begin{eqnarray} \label{eq:MixedBC} \notag
\Pi_- \phi|_{x_3=0} &=& 0 \; , \\ \notag
\partial_3\Pi_+ \phi|_{x_3=0} &=& 0 \; , \\
\Pi_- + \Pi_+ &=& 1 \;.
\end{eqnarray}
The Green's function satisfying above conditions can be found by using the method of images

\begin{figure}[h]
\begin{center}
  \includegraphics[width=80mm]{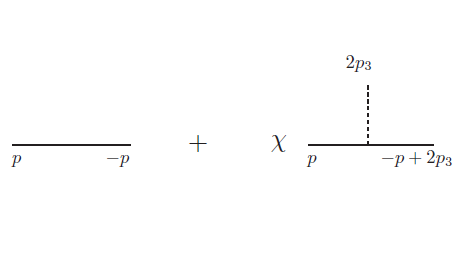}
  \caption{\small Scalar propagator for mixed boundary conditions. Dashed line represents the momentum reflected from the boundary. In this notation $p_3=(\underline{0},p_3)$.}
  \label{fig:P1}
  \end{center}
\end{figure}
\begin{equation} \label{eq:Scalar2point}
\vev{\phi(x) \phi(x')}= C_d \left( \frac{\ONE}{|x-x'|^{(d-2)}} +  \frac{\chi}{|\tilde{x}-x'|^{(d-2)}} \right) \quad ,
\end{equation}
where $\tilde{x}= (\underline{x},-x_3)$ is the position of the image charge, $\chi= \Pi_+ - \Pi_-$ (the scalar indices have been suppressed) and $C_d= \frac{\Gamma(\frac{d}{2})}{2 (d-2) \pi^{\frac{d}{2}}}$. We will now analytically extend this function to the upper semi-space $x_3, x_3' \geq 0$ and Fourier transform it. Doing this we will obtain a propagator depicted on Figure \ref{fig:P1}. The first term is the standard propagator $\frac{1}{p^2}$, with momentum being conserved across. The second term is new- it is also proportional to $\frac{1}{p^2}$ but now the perpendicular momentum is not being conserved, which is expressed by the dashed line. This extra term can be intuitively understood as an elastic reflection from the planar boundary. \\ 

\subsection{Perturbation theory with boundary} \label{sec:Feynman}
\begin{figure}[h]
\begin{center}
  \includegraphics[width=80mm]{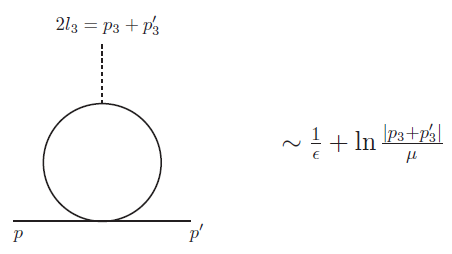}
  \caption{\small A boundary contribution to the scalar two-point function with $\phi^4$ interaction.}
  \label{fig:P3}
  \end{center}
\end{figure}

\begin{figure}[h]
\begin{center}
  \includegraphics[width=80mm]{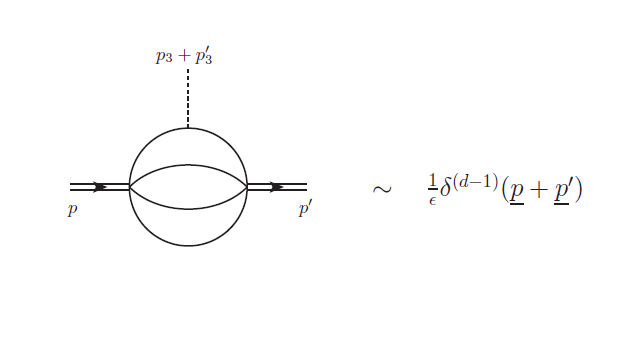}
  \caption{\small A boundary contribution to the correlator of $\phi^4$ in momentum space.}
  \label{fig:P4}
  \end{center}
\end{figure}

For any given Feynman diagram we can treat the second term on Figure \ref{fig:P1} as the usual propagator and the dashed line as an external momentum emission (even if it is attached to virtual propagators) as long as we include the overall delta function
\begin{equation} \label{eq:deltaP3}
\delta (\sum_{i} p_3^{i}) \quad ,
\end{equation}
where the sum goes over perpendicular components of \textit{all} external momenta, \textit{including} the dashed lines. Including bulk vertices is possible provided they are invariant w.r.t. parity-like transformation $\phi(\underline{x},-x_3) \to \chi \phi(\underline{x},x_3)$, where $\chi$ was defined under \eqref{eq:Scalar2point}.\footnote{As an example one can take single scalar with $\lambda \phi^4$ since under the reflection it changes as $\lambda \to \chi^4 \lambda= \lambda$. More generally $\lambda$ has to be $\chi-$invariant tensor.} In this case each vertex has to be multiplied by a factor $\frac{1}{2}$ to avoid the double counting that arises from including the $x_3>0$ contribution, but otherwise can be treated as usual conserving all components of the incoming momentum.\footnote{We can also include boundary interactions, that only conserve parallel momentum by using that $\delta(x_3)= \frac{1}{2 \pi}\int_{-\infty}^{\infty} d L_3 e^{iL_3 x_3}$, so that the relevant vertex will be proportional to $\int d L_3 e^{iL_3 x_3}$ with $L_3= \sum_{incoming} p_3^i$. }
These rules allow us for systematic treatment of perturbative computations in scalar field theory with boundary. The same rules apply to theory with fermions, the only thing that changes is the form of free propagator and $\chi$. Note that the presence of delta function depending on internal momenta \eqref{eq:deltaP3} allows for some new phenomena. In particular would-be power divergences now become logarithmic and we need to include new counterterms to renormalize those. Such (local) counterterms will exist purely on the boundary and will lead to some additional renormalization. \\
The main idea can be illustrated on the free-scalar example at hand by adding  $\phi^4$ interaction (see Figure \ref{fig:P3}).
The total delta function \eqref{eq:deltaP3} imposes $2l_3=p_3+p_3'$ for the perpendicular component of loop momentum and we are left with $d^{(d-1)} \underline{l}$ loop integral. The remaining integral will be nonzero and equal the $2d$ massive bubble diagram with a mass $m= \frac{1}{2}(p_3+p_3')$ and thus needs to be renormalized by adding a (divergent) `boundary mass' counterterm $\frac{1}{\epsilon}\int d^3 x \delta(x_3) \phi^2(x)$. Thus we see that bulk interactions induce renormalization of purely boundary terms. The same conclusion was reached in the original paper \cite{Symanzik:1981wd} and more recently renormalization of boundary couplings in $d=4- 2\eps$ was also discussed in \cite{Herzog:2017xha}. For more detailed analysis including the discussion of renormalizability we refer the reader to \cite{Diehl1981}. \\
Next we would like to discuss the renormalization of composite operators. To this end we can use the local coupling formalism, where the coupling to a composite operator $O$ gets promoted to a function of space
\begin{equation}
\int d^d x \lambda O(x)  \to \int d^d x \lambda(x) O(x) \quad.
\end{equation}
The correlators of $O$ are then obtained by taking functional derivatives of the partition function w.r.t. $\lambda(x)$. In \cite{Jack:1990eb} it was shown that doing this it is necessary to include additional divergent counterterms dependent on $\lambda(x)$ and its derivatives. These have to be include in order to cancel UV- divergent contact terms in the composite operator correlators. When including the planar boundary we expect some counterterms proportional to $\delta(x_3)$ to appear. Indeed, for example by considering again $O= \phi^4$ one observes that the diagram on Figure \ref{fig:P4} in $d=3-2\epsilon$ now includes logarithmic divergence (a pole in $\epsilon$).
Once again we need to include, purely boundary term $\frac{1}{\epsilon}\int d^d x \delta(x_3) \lambda(x)^2$ to cancel this divergence. A detailed local coupling analysis of scalar field theories with boundaries was given in \cite{McAvity:1992fq}. \\
In the context of this paper we treat the metric $g_{\mu \nu}$ as the local coupling to the energy-momentum tensor. The relevant counterterms are then the ones invariant under boundary diffeomorphisms  \eqref{eq:bulkCterms}. 

\subsection{Energy-momentum tensor correlator} \label{sec:EMT-scalar}
\begin{figure}[h]
\begin{center}
  \includegraphics[width=80mm]{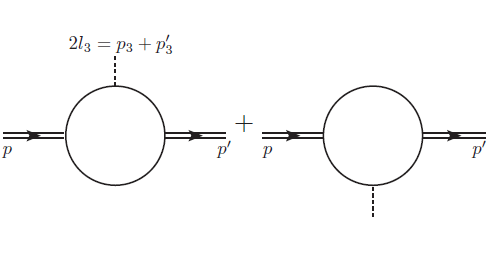}
  \caption{\small The boundary contribution to the two-point function of $T_{\mu \nu}$ (inserted at the sides).}
  \label{fig:P2}
  \end{center}
\end{figure}

The relevant (bare) two-point functions are defined through the variational principle \eqref{eq:QAPTT}
with the rule \eqref{eq:BoundVariation} assumed
\begin{equation} \label{eq:TT0def}
\vev{ T_{ij}(x) T_{kl}(x')}= -4\vev{ \frac{\delta S_0}{\delta g_{ij}(x)}  \frac{\delta S_0}{\delta g_{kl}(x')} } \big |_{g_{ij}=g_{kl}=\delta_{ij}}+ 4\vev{ \frac{\delta^2 S_0}{\delta g_{ij}(x)\delta g_{kl}(x')}}\big |_{g_{\mu \nu}= \delta_{\mu \nu}} \;.
\end{equation}
The first term on r.h.s. of \eqref{eq:TT0def} is defined is simply the correlator of \eqref{eq:scalarEMT} at non-coincident points. In momentum space it corresponds to the two diagrams on Figure \ref{fig:P2}. The other possible diagrams include the usual $d=3$ contribution with no dashed outgoing lines and the other `reflected' diagram with two outgoing dashed lines. The latter two diagrams are proportional to $\delta(p_3+p_3')$ and $\delta(p_3-p_3')$ respectively and both of these distributions vanish for the momentum configuration of \eqref{eq:TTdivApp}. The nonvanishing diagrams of Figure \ref{fig:P2} correspond to absorption of virtual scalar by the boundary and can be computed via standard methods (we refer the reader to the Appendix \ref{app:A}, where some more details of these computations are given). \\
 The second piece in \eqref{eq:TT0def} is a contact term. In fact, as we saw in Section \ref{sec:SchemInv}, no purely background counterterms contribute to \eqref{eq:TT0def} in $d=3$ so the only possible contribution comes from varying the second term in \eqref{eq:S0scalar}. This contribution is new compared to the boundary-less case where $\vev{\phi^2}= 0$. We have
\begin{equation} \label{eq:SemiLocalScalar}
\vev{ \frac{\delta^2 S_0}{\delta g_{ij}(x)\delta g_{kl}(x')}}\big |_{g_{\mu \nu} = \delta_{\mu \nu}} =
\frac{1}{2} \xi \vev{\phi^2(x)} D_{ijkl} \left(\delta^{(3)}(x-x') + \delta^{(3)}(\tilde{x}-x') \right) \; ,
\end{equation}
where $D_{ijkl}$ is a differential operator quadratic in derivatives that is obtained from the variation $\frac{\delta^2}{\delta g_{ij}\delta g_{kl}}\int d^3x  \sqrt{g} R$. This operator is evanescent near $d=3$, which will leave a finite imprint when combined with the divergent Fourier transform of $\vev{\phi^2(x)} \sim \Tr \chi\frac{1}{|x_3|}$ in \eqref{eq:SemiLocalScalar}. Notice that this is an example of `mean field' contribution discussed around \eqref{eq:MFImprov}.\\
We are interested in the traceless operator $\tilde{T}_{ij}$ defined in \eqref{eq:TracelessT} 
\begin{equation}
\tilde{T}_{ij}= T_{ij}- \frac{1}{(d-1)} \delta_{ij} T_{kk}= \partial_i \phi \partial_j \phi - \frac{1}{(d-1)} \delta_{ij} (\partial_k \phi)^2+ \xi ( \frac{1}{(d-1)}  \delta_{ij} \partial_k^2  - \partial_i \partial_j) \phi^2 \quad.
\end{equation}
Correlators of $\tilde{T}_{ij}$ are constructed from the traceless projection of \eqref{eq:TT0def} so we now have all the pieces needed for the computation of \eqref{eq:TilATen}. Evaluating the diagrams on Figure \ref{fig:P2} and including the Fourier transform of \eqref{eq:SemiLocalScalar} (see Appendix \ref{app:A}) we finally arrive at
\begin{equation} \label{eq:scalarConsTT}
p_i \vev{\tilde{T}_{ij}(\underline{p},p_3) \tilde{T}_{kl}(\underline{p'},0)}= -\Tr \chi \frac{1}{4 \pi}(\frac{1}{6}- \xi) p_j \left( p_k p_l - \frac{1}{(d-1)} \delta_{kl} p_m^2 \right) \delta^{(d-1)}(\underline{p}+ \underline{p}') + \mathcal{O} \left(\frac{\underline{p}^4}{p_3^2}\right) ,
\end{equation}
which yields the central charge
\begin{equation}
c_\partial= -\frac{1}{4 \pi} \Tr (\chi) \left( \frac{1}{24}- \frac{1}{4} \xi \right) ,
\end{equation}
where we used \eqref{eq:TheRelation}. The trace anomaly follows from \eqref{eq:TraceAnBoundary} and we find agreement with the result of \cite{Vassilevich:2003xt} (including the dependence on $\xi$ consistent with \eqref{eq:ImprovContr}). For conformal theory we additionally require $\xi_c= \frac{1}{8}$ and get 
\begin{equation} \label{eq:FreeScFinal}
c_\partial= -\frac{1}{4 \pi} \frac{1}{96} \Tr (\chi)  \;.
\end{equation}
For single scalar field the trace is $\Tr (\chi) =\pm 1$ with positive sign for Neumann and negative sign for Dirichlet boundary conditions. In this case \eqref{eq:FreeScFinal} reduces to the result of \cite{Jensen:2015swa} obtained by computing the partition function of conformally coupled scalar on a hemisphere. We also computed the anomaly of free fermions by the same method and verified that it vanishes  as seen in the literature \cite{Fursaev:2016inw}.

\section{Conclusions and outlook}

The main goal and result of this paper was to derive the relations \eqref{eq:TheRelation}, \eqref{eq:SumRuleMain} between the central charge $c_\partial$ and flat space EMT correlators \eqref{eq:TilATen}. We saw that the bulk contributes local divergent terms \eqref{eq:Wbound} to the generating functional $W$ at the boundary, which are then related to a finite contact term of \eqref{eq:TilTil1} in the small $\frac{|\underline{p}|}{p_3}$ regime. \\
This situation is similar to integrating out a massive particle in a $2d$ (cf. \eqref{eq:2dMassTT} with $c_{\text{\tiny{IR}}}=0$ and $c_{\text{\tiny{UV}}}=c$), which leaves behind a local term in the IR effective action proportional to the central charge (such setup can be realized by coupling the theory to a background dilaton and computing the resulting IR effective action \cite{Komargodski:2011xv}). In our case the the perpendicular momentum $p_3$ played the role of mass parameter when integrating out the bulk state to obtain $c_\partial$. Physically the relevant amplitude \eqref{eq:TTcorrLowMom} corresponds to the process where the incoming bulk state is being absorbed by the boundary.  \\
  As a side result we obtained a criterion \eqref{eq:ConsistencyCondCC} guaranteeing no boundary divergences in the two point function of perpendicular components $\vev{T_{33}(\underline{p},p_3),T_{33}(\underline{p},0)}$, which is consistent with the examples in the literature. \\
 The present method only allows us to find the bulk contribution to $c_\partial$ coming from correlators of $T_{\mu \nu}$ coupled to a bulk metric with boundary conditions \eqref{eq:MetricBC}. 
 It would be interesting to extend the analysis of this paper to include the computation of $c_\partial$ in presence of non-trivial boundary operators (for example in theories that have non-vanishing $O_{ij}$ with bulk-boundary interactions). Since $\Delta c_\partial$ under boundary RG flow only depends on the dynamics of boundary operators \cite{Jensen:2015swa}, it would not be unreasonable to expect that the bulk and boundary contributions simply sum. \\
 The second main contribution of this paper is development of a general perturbative framework in Section \ref{sec:Feynman}, where the boundary can be treated as an external state (see for example Figure \ref{fig:P2}). Momentum space methods have already been applied to boundary field theories in the literature \cite{Diehl1981, Symanzik:1981wd, Herzog:2017xha}. Where this work complements the previous analysis is in including also the Fourier transform in the \textit{non-conserved} direction. We saw that doing this allows one to use standard diagrammatic techniques and therefore can be applied in a broad context. A natural extension of this work would be to check the method in examples where the bulk is weakly interacting and compute the perturbative corrections to $c_\partial$. \\

\appendix
\numberwithin{equation}{section}

\subsection*{Acknowledgements}

The author is thankful to Zohar Komargodski for critical remarks and ongoing fruitful discussions. The author is also grateful to Guido Festuccia and Adam Schwimmer for feedback on the final draft and Agnese Bissi, Tobias Hansen, Marjorie Schillo and Alexander Söderberg for useful discussions.  
The author is supported by the ERC STG grant 639220 (curvedsusy).

\section{Derivation of the central charge relation} \label{app:C}

The expansion \eqref{eq:LowMomOO} applied to $T_{ij}$ will have the form
\begin{equation} \label{eq:TTcorrLowMom}
\vev{T_{ij}(\underline{p},p_3)T_{kl}(-\underline{p},0)} \approx \left(A_{ijkl}(\underline{p})+ p_3^2 B_{ijkl} \right) \delta^{(d-1)} (\underline{p}+\underline{p}') + \mathcal{O} \left(\frac{\underline{p}^4}{p_3^2}\right) \;.
\end{equation}
It would be tempting now to equate $A_{ijkl}$ with $T_{ijkl}$ from \eqref{eq:Tan}. Such argument is complicated by the fact that $T_{ij}$ is not conserved in the parallel direction, which means that non-covariant (finite) contact terms also contribute to $A_{ijkl}$. In fact, the general form of such correction is
\begin{equation} \label{eq:divAcontribution}
p_i A_{ijkl} \sim  p_j (p_k p_l - \delta_{kl} \underline{p}^2) \; ,
\end{equation}
which follows from the transverse property $p_k T_{kl}(-\underline{p},0)= 0$ valid at $p_3'=0$.\footnote{Note that this identity is only valid on-shell, however in this case the equations of motion only contribute to the $p_3^2$ component} 
We verified that \eqref{eq:divAcontribution} appears already in the free scalar example. The anomaly \eqref{eq:Tan} should be equal to the trasverse component of $A_{ijkl}$ so the terms contributing to \eqref{eq:divAcontribution} need to be projected away.
We can either do this on a case-by-case basis via shifting $A_{ijkl}$ by a contact term proportional to 
\begin{equation}\label{eq:divAcontributionCT}
\delta_{ij} (p_k p_l - \delta_{kl} \underline{p}^2)
\end{equation}
 or we can work with correlators of the traceless component \eqref{eq:TracelessT}. The traceless correlator is invariant under the shift \eqref{eq:divAcontributionCT} and therefore the terms \eqref{eq:divAcontribution} don't contribute to it. \\
As a side note, let us comment that it can be shown by using formulas of \cite{McAvity:1993ue} that for $|\underline{x}-\underline{x'}| \gg |x_3|$ the two-point function of $\tilde{T}_{ij}$ behaves as
\begin{eqnarray} \label{eq:IRconserved}
(\underline{x}-\underline{x'})^2 \vev{\tilde{T}_{ij}(\underline{x},x_3)\tilde{T}_{kl}(\underline{x}',x_3')} \to \frac{\epsilon(1)}{\underline{s}^{2(d-1)}}\left(\delta_{ij} \delta_{kl}-(I_{ik}(\underline{s}) I_{jl}(\underline{s}) +I_{il}(\underline{s}) I_{jk}(\underline{s}) ) \right)  \;,
\end{eqnarray}
where $\underline{s}= \underline{x}-\underline{x'}$ and $I_{ij}(\underline{x})= \delta_{ij}- \frac{2 x_i x_j}{\underline{x}^2}$. The constant $\epsilon(1)$ is a related to a specific kinematic structure defined in \cite{McAvity:1993ue} and it depends on boundary conditions. For $d=3$ the r.h.s. of \eqref{eq:IRconserved} looks exactly like the 2-point function of conserved EMT in a $2d$ CFT.\footnote{The extra scale factor $(\underline{x}-\underline{x'})^2$ on the l.h.s. of \eqref{eq:IRconserved} cancels the one from Fourier integrals over $dx_3, dx_3'$.}\\
Before we proceed, let us discuss some further desirable properties of the operator $\tilde{T}_{ij}$. We saw at the end of Section \ref{sec:BoundaryCT} that the two-point functions of $T_{\mu \nu}$ receive non-trivial contributions from equations of motion in the presence of boundary. One possible source of e.o.m.  ambiguity\footnote{Here we assume a symmetric energy-momentum tensor. In general there are also ambiguities related to Lorenz transformations, which induce antisymmetric improvements.}  comes from defining dimensions of fundamental fields under conformal (Weyl) transformations \cite{Fujikawa:1980rc}
\begin{equation}
T_{\mu \nu} \to T_{\mu \nu} + c \delta_{\mu \nu} \phi(x) \frac{\delta S}{\delta \phi(x)} \quad.
\end{equation}
Such ambiguities leave \eqref{eq:TracelessT} invariant. \\
Next we study the bulk operator improvement of the form
\begin{eqnarray} \label{eq:ImprovementGeneral}
\delta S =\frac{1}{2} \int d^d x \sqrt{g} \xi R O 
\end{eqnarray}
for some local operator $O$ of scaling dimension $(d-2)$. The contribution of such improvement is two-fold. First, one has to shift the energy-momentum tensor by $(\partial_\mu \partial_\nu - \Box \delta_{\mu \nu})O$, which changes the correlators of $T_{\mu \nu}$ at non-coincident points.  The leading term in the expansion \eqref{eq:TilTil1} doesn't change under such shift. This can be shown by noting that the relevant contribution of the above shift to $\vev{\tilde{T}_{ij}\tilde{T}_{kl}}$ will always be of the form $\underline{p}^2\vev{O \tilde{T}_{ij}}$. From the tracelessness of $\tilde{T}_{ij}$, dimensional analysis and $SO(2)$ covariance one infers that $\vev{O \tilde{T}_{ij}} \sim 0 + \mathcal{O}\left(\frac{\underline{p}^2}{p_3^2}\right)$ so the improvement only contributes at $\mathcal{O}\left(\frac{\underline{p}^4}{p_3^2}\right)$.\\
 Second, as can be seen in the free scalar example, the $\xi$-dependence will still come in through Fourier transforming the contact term of the type \eqref{eq:SemiLocalContact} obtained by varying  \eqref{eq:ImprovementGeneral} w.r.t. metric twice. Alternatively it can be seen as a mean-field contribution to the divergent part of the effective action \eqref{eq:Wbound}
\begin{eqnarray} \label{eq:MFImprov}
\delta W_{an} = \vev{\delta S} = \frac{1}{2}\int d^d x \sqrt{g} \xi R \vev{O} \quad.
\end{eqnarray}
Close to the boundary $\vev{O} \sim \frac{1}{|x_3|^{d-2}}$, so the above integral is exactly of the form \eqref{eq:Wbulk} and therefore contributes to $c_\partial$. \\
We will now proceed to identify the central charge in the the low $\underline{p}^2$ expansion of the two-point function \eqref{eq:TilTil1}. From \eqref{eq:divAcontribution} we can deduce\footnote{In more detail the identity \eqref{eq:TilATen} follows by writing 
\begin{equation*}
p_i \vev{\tilde{T}_{ij} \tilde{T}_{kl}}= p_i \vev{T_{ij} \tilde{T}_{kl}}- \frac{1}{2} p_j \vev{T_{mm} \tilde{T}_{kl}}\; .
\end{equation*}
When expanding the above in $\underline{p}^2$, the leading contribution coming from both of the terms on the right-hand side will be proportional to $p_j \left(p_k p_l - \frac{1}{(d-1)} \delta_{kl} \underline{p}^2\right)$. For the first term this follows by taking the traceless part of \eqref{eq:divAcontribution} and for the second by writing down the relevant $SO(2)-$covariant expression traceless in $k,l$.} 
\begin{eqnarray} \label{eq:TilATen}
p_i \tilde{A}_{ijkl} =  \kappa p_j \left(p_k p_l - \frac{1}{(d-1)} \delta_{kl} \underline{p}^2\right) \; ,
\end{eqnarray}
where $\kappa$ is a model-dependent constant.\footnote{One slightly confusing point here is that although we have argued that the non-covariant terms don't contribute to the traceless part $\tilde{A}_{ijkl}$, the relation \eqref{eq:TilATen} implies its non-transverse character. The reason for this apparent discrepancy is the fact that traceless part of transverse projector is itself not transverse in $d>1$.} Due to presence of $p_3>0$, which acts as an effective IR cutoff, the small $\underline{p}$ behaviour of \eqref{eq:TilTil1} should be under control and as previously seen all correlators of $T_{ij}$ are scheme-independent and finite so $\kappa$ is a well defined and unambiguous quantity.\footnote{The scale invariance prevents the $\kappa$ from depending on $p_3$ in strictly $d=3$.} \\ 
On the other hand from \eqref{eq:QAPTT} and \eqref{eq:Wbound} it is seen that  \eqref{eq:TilATen} should correspond to the traceless part of \eqref{eq:Tan}. The traceless part of \eqref{eq:Tan1} satisfies
\begin{eqnarray} \label{eq:divApos}
p_i \tilde{T}_{ijkl}= -4 c_\partial pj \left(p_k p_l - \frac{1}{(d-1)} \delta_{kl} \underline{p}^2\right) \;.
\end{eqnarray}
By equating \eqref{eq:divApos} with \eqref{eq:TilATen} we finally arrive at
\begin{equation}\label{eq:TheRelation1}
\kappa= -4 c_\partial \quad.
\end{equation}

\section{Explicit computation of momentum space integrals} \label{app:A}
 
In this appendix we would like to flesh out the one-loop computation of the scalar amplitude
\begin{eqnarray} \label{eq:TTdivApp}
p_i \vev{\tilde{T}_{ij}(\underline{p},p_3)\tilde{T}_{kl}(-\underline{p},0)}
\end{eqnarray}
that we used in Section \ref{sec:EMT-scalar}. It turns out, that the only the diagrams of Figure \ref{fig:P2} give nonzero contribution to this amplitude.
To compute the momentum space correlators we proceed as in \cite{Hathrell:1981zb} and start by writing down the vertex 
\begin{eqnarray} \label{eq:VijScalar}
\tilde{V}_{ij}(q_1,q_2)= q_{1 i} q_{2 j} + q_{1 i} q_{2 j}- \frac{2}{(d-1)}\delta_{ij} (\underline{q_1}\cdot \underline{q_2})
\end{eqnarray}
corresponding to the insertion of \eqref{eq:scalarEMT} with $\xi=0$ and conserved parallel momentum $\underline{p}= \underline{q_1}+\underline{q_2}$ for two incoming scalars with momenta $q_1, q_2$.  In that case it can be shown that
\begin{equation}
    p_i\tilde{V}_{ij}=\frac{p_j (d-3)}{(d-1)}(\frac{1}{2}\underline{p}^2-\frac{1}{2}\underline{q}_1^2-\frac{1}{2}\underline{q_2}^2)+ \underline{q}_1^2 q_{2 j}+\underline{q}_2^2 q_{1 j} \quad.
\end{equation}
Using this identity helps to simplify the tensor structures in the diagrams on Figure \ref{fig:P2} leaving us with the only non-vanishing contribution 
\begin{eqnarray} \label{eq:TTdivergenceMom}
-2 \frac{1}{2} \Tr(\chi) p_j (\frac{1}{2}p_3)^2 \int \frac{d^d l}{(2\pi)^{d-1}} \delta(2l_3+p_3) \frac{l_k (l-p)_l+ l_l(l-p)_k-\frac{2 \delta_{kl}}{(d-1)} \underline{l}\cdot (\underline{l}-\underline{p})}{l^2(l-p)^2} \quad ,
\end{eqnarray}
where we the first factor of $2$ comes from counting in the both diagrams Figure \ref{fig:P2} which give identical contributions. Imposing the delta function constraint we end up with with a massive one-loop diagram of 'effective mass' $m= \frac{1}{2}p_3$ in $d=2-2\eps$. After the Feynman parametrisation the divergent piece of this integral cancels out and \eqref{eq:TTdivergenceMom} becomes
\begin{eqnarray}
  \frac{1}{4 \pi} \Tr(\chi) m^2 p_j \left(p_k p_l - \frac{1}{(d-1)} \delta_{kl} \underline{p}^2 \right)\int_{0}^1 \frac{x(1-x) dx}{x(1-x)\underline{p}^2+m^2} \quad.
\end{eqnarray}
Due to the presence of non-zero mass the above integral has regular behaviour at small $\underline{p}^2$ and we are able to get the final result 
\begin{equation}
   \frac{1}{6} \Tr(\chi) \frac{1}{4 \pi} p_j \left(p_k p_l - \frac{1}{(d-1)} \delta_{kl} \underline{p}^2 \right) + \mathcal{O}(\frac{\underline{p}^4}{p_3^2}) \quad.  
\end{equation}
Extension to $\xi \neq 0$ is easy, since it turns out that the contributions proportional to $\xi$ in the diagrams on Figure \ref{fig:P2} only contribute at $\mathcal{O}(\frac{\underline{p}^4}{p_3^2})$ and thus we can ignore them for the purposes of this paper. The contribution from the second term in \eqref{eq:TT0def} is obtained by projecting \eqref{eq:SemiLocalScalar} to its traceless component $\tilde{D}_{ijkl}$, which satisfies
\begin{eqnarray}
\partial_i \tilde{D}_{ijkl}= -\frac{2(d-3)}{(d-1)}\partial_j \left(\partial_k\partial_l - \frac{1}{(d-1)}\delta_{kl}  \underline{\partial}^2 \right).
\end{eqnarray}
We can use this to evaluate the Fourier transform of 
\begin{equation} \label{eq:SemiLocalScalar2}
\frac{1}{2}\xi \vev{\phi^2(x)} \partial_i \tilde{D}_{ijkl} \left(\delta^{(3)}(x-x') + \delta^{(3)}(\tilde{x}-x') \right) \;.
\end{equation}
Using that $\vev{\phi^2}= \frac{\Tr(\chi)}{4 \pi} \frac{1}{2|x_3|}$ and computing the Fourier integrals over parallel coordinates and $dx_3'$ (with $p_3'=0$) we reduce the Fourier integrals to one contribution
\begin{equation}
2 \xi p_j \frac{(d-3)}{2(d-1)} \frac{1}{4 \pi} \Tr(\chi) p_j \left(p_k p_l - \frac{1}{(d-1)}\delta_{kl} \underline{p}^2 \right) \int dx_3  \frac{e^{ip_3 x_3}}{|x_3|^{(d-2)}} \quad.
\end{equation}
After performing the last Fourier transform we are left with a finite contribution
\begin{eqnarray}
-\Tr(\chi) \xi \frac{1}{4 \pi} p_j \left(p_k p_l - \frac{1}{(d-1)}\delta_{kl} \underline{p}^2 \right) + \mathcal{O}(\epsilon) \;.
\end{eqnarray}

\section{Divergent contribution to $\vev{T_{33}T_{33}}$} \label{app:B}
In this appendix we would like to illustrate how divergences arise in the correlator
$\vev{T_{33}(\underline{p},p_3)T_{33}(-\underline{p},0)}$ arise. The momentum space vertex for $T_{33} \phi \phi$ reads
\begin{eqnarray} \label{eq:V33vertex}
V_{33}(q_1, q_2)= (q_1)_3 (q_2)_3 - (\underline{q_1} \cdot \underline{q_1})+ 2\xi \underline{p}^2 \;.
\end{eqnarray}
The relevant amplitude is obtained from one-loop diagram on Figure \ref{fig:P4}. Just as in Appendix \ref{app:A}, the integration over perpendicular loop momentum is removed by $\delta(2l_3 - p_3)$ and we are left with $d^{(d-1)}\underline{l}$ loop integral with effective mass $m=\frac{1}{2}p_3$. The numerator of the integrand will contain four powers  of momenta arising from two vertices \eqref{eq:V33vertex} and we can expand it in terms of inverse powers of propagators $P_1=\underline{l}^2 + m^2 , \; P_2= (\underline{p}- \underline{l})^2+ m^2 $. By doing this we find two divergent pieces. The first one proportional to $p_3^2$
\begin{eqnarray}
-\frac{1}{8} \Tr \chi p_3^2 \int \frac{d^{d-1} \underline{l}}{(2 \pi)^{(d-1)}}\frac{P_1+P_2}{P_1 P_2}= \frac{1}{4 \eps}\frac{1}{4 \pi} \;.
\end{eqnarray}
This divergence cancels against the contribution (cf. \eqref{eq:TTCorrDef}) from the Fourier transform of
\begin{eqnarray}
\frac{\delta^2 S_0}{\delta g_{33}(x) \delta g_{33}(x')}  \sim \left( \delta^{(d)}(x-x')+ \delta^{(d)}(x+x') \right) \frac{\Tr \chi}{|x_3|^{3-2\eps}} \; ,
\end{eqnarray}
where \eqref{eq:BoundVariation} was used to extend the delta function to the upper half-space. The cancellation is expected in general since one can check by inspection that there are no boundary diffeomorphism-invariant  counterterms involving $g_{33} \partial_3^2 g_{33}$, which means that any potential $p_3^2$ divergences in $\vev{T_{33} T_{33}}$ have to be cancelled by the contribution from $\vev{ \frac{\delta^2 S_{dyn}}{\delta g_{33}(x)\delta g_{33}(x')}}\big |_{g_{\mu \nu}\to \delta_{\mu \nu}}$. \\
 The second divergent piece reads
\begin{eqnarray}
\frac{1}{4 \pi} \frac{2}{\eps} \Tr \chi \left(\xi-\frac{1}{8}\right)\underline{p}^2 \quad.
\end{eqnarray}
This term has to be cancelled by adding a divergent counterterm \eqref{eq:RndoubleVariation}. Note however, that for conformally coupled scalar with $\xi_c=\frac{1}{8}$ this contributions vanishes on its own as expected from the condition \eqref{eq:ConsistencyCondCC}. The cancellation of divergences is consistent with the results of \cite{McAvity:1993ue}, where the boundary limit of $\vev{T_{33}T_{33}}$ ($\alpha(v) \to \alpha(1)$ in their language) for conformally coupled scalar contains no divergences and also with the explicit heat kernel computation of \cite{Branson:1990xp}.

\bibliographystyle{utphys}
\bibliography{BoundaryMomentumFinal}

\end{document}